\documentclass[final,3p,times]{elsarticle}

\usepackage{amssymb}

\usepackage{hyperref}
\usepackage{amsmath}
\usepackage{graphicx}
\usepackage{color}
\usepackage{comment}
\usepackage{lineno}

\newcommand{\red}[1]{\textcolor{red}{#1}} 
\newcommand{\pr}[1]{\left(#1\right)}

\newcommand{\lat}[1]{$(#1)$}
\newcommand{\unit}[1]{~\mathrm{#1}}

\newif\ifAPPEND
\APPENDtrue

\renewcommand{\red}[1]{\iffalse #1 \fi} 

\newif\ifENG
\newif\ifJPN
\ENGtrue   \JPNfalse     

\journal{Nuclear Inst. and Methods in Physics Research, A}
\begin{document}
\begin{frontmatter}



\title{Demonstration of Near-Epithermal Neutron Reflective Optics}

\author[Nagoya,RIKEN]{Takuhiro~Fujiie}
\ead{fujiie@phi.phys.nagoya-u.ac.jp}
\author[Nagoya]{Ryota~Abe}
\author[KUR]{Masahiro~Hino}
\author[Nagoya]{Mayu~Hishida}
\author[RIKEN]{Takuya~Hosobata}
\author[KMI,Nagoya]{Masaaki~Kitaguchi}
\author[Nagoya]{Rintaro~Nakabe}
\author[J-PARC]{Kenichi~Oikawa}
\author[Nagoya]{Takuya~Okudaira}
\author[CROSS]{Joseph~D.~Parker}
\author[J-PARC]{Kenji~Sakai}
\author[Nagoya]{Hirohiko~M.~Shimizu}
\author[J-PARC]{Yusuke~Tsuchikawa}
\author[RIKEN]{Yutaka~Yamagata}

\affiliation[Nagoya]{organization={Dept. of Phys. Nagoya University},
            addressline={Furocho Chikusa},
            city={Nagoya},
            postcode={464-8602},
            state={Aichi},
            country={Japan}}

\affiliation[RIKEN]{organization={RIKEN Center for Advanced Photonics},
            addressline={Hirosawa 2-1},
            city={Wako},
            postcode={351-0198},
            state={Saitama},
            country={Japan}}

\affiliation[KUR]{organization={Institute for Integrated Radiation and Nuclear Science, Kyoto University},
            addressline={2, Asashiro-Nishi, Kumatori},
            city={Sennan-gun},
            postcode={590-0494},
            state={Osaka},
            country={Japan}}

\affiliation[KMI]{organization={Kobayashi-Maskawa Institute,
            Nagoya University},
            addressline={Furocho Chikusa},
            city={Nagoya},
            postcode={464-8602},
            state={Aichi},
            country={Japan}}
            
\affiliation[J-PARC]{organization={J-PARC Center, Japan Atomic Energy Agency (JAEA)},
            addressline={2-4 Shirakata},
            city={Tokai},
            postcode={319-1195},
            state={Ibaraki},
            country={Japan}}
\affiliation[CROSS]{organization={Comprehensive Research Organization for Science and Society (CROSS) Neutron Science and Technology Center},
            addressline={162-1 Shirakata},
            city={Tokai},
            postcode={319-1106},
            state={Ibaraki},
            country={Japan}}

\begin{abstract}

Specular reflection of neutrons on material surfaces has been demonstrated in the energy range of $0.09$--$0.7~\unit{eV}$. 
The results suggest that the applicable energy range of reflective neutron optics can be extended to the near-epithermal region by using existing techniques.

\end{abstract}




\end{frontmatter}



\section{Introduction}

The successful operation of spallation neutron sources, driven by intense proton beams, has enabled energy-resolved intense neutrons in the thermal and cold regions and also in the epithermal regions\cite{Mason2006-ik, Wei2009-ae, Nagamiya2012-xa, Nakajima2017-hb}.
In this paper, we refer to neutrons with kinetic energy $E_{\rm n}$ in the region of 0.5--1 eV as ``near-epithermal'' neutrons.
The energy resolution of the pulsed near-epithermal neutron beam facilitates efficient observations of neutron scattering with a large energy transfer in materials and precise studies of neutron-induced compound nuclear states for symmetry breaking at fundamental levels\cite{Okudaira2018-du, Flambaum2019-pt, Snow2021-jn}.

The use of transport optics for near-epithermal neutrons can significantly improve the efficiency of neutron applications in fundamental and practical experiments.
A specular neutron reflector on the surface and interface of materials is the most commonly utilized optical device, which transports grazing-angle incident neutrons\cite{Utsuro2010-oh}. 
The maximum grazing angle of the reflection $\theta_{\rm max}$ is given as the ratio of the normal component of neutron momentum to $\hbar k_{\rm max} = (2 m_{\rm n} U)^{1/2}$, where $m_{\rm n}$ represents the neutron mass and $U$ is the Fermi pseudopotential\cite{Fermi1936-hh}.
Additional reflectivity beyond $\theta_{\rm max}$ can be introduced by a multilayer coating known as a supermirror.
The maximum grazing angle of the additional reflectivity is commonly measured as the ratio of $\theta_{\rm max}$ to the natural abundance of nickel, which is referred to as the $m$-value of the supermirrors\cite{Hino2009-kp}.
The $m$-value of supermirrors, commonly used for cold and thermal neutron transport, is typically within the range of 2--3\cite{Abele2006-sq, Mishima2009-yx, Mishima2015-nm, Ebisawa1979-nb, Ebisawa1995-ep, Kawabata2002-wz}.
However, near-epithermal neutron transport requires the $m$-value of approximately 10 or greater.
In addition, the shape of the mirror requires 10 times more precision than that of conventional mirrors.
For these reasons, specular reflection of near-epithermal neutrons has not been observed.
In this paper, we report on an experimental demonstration of specular reflection of near-epithermal neutrons on a flat material surface as the first step towards the practical application of near-epithermal reflective optics.


\section{Experimental Geometry}

\begin{figure*}[htb]
\centering
 \includegraphics[keepaspectratio, width=0.5\linewidth]{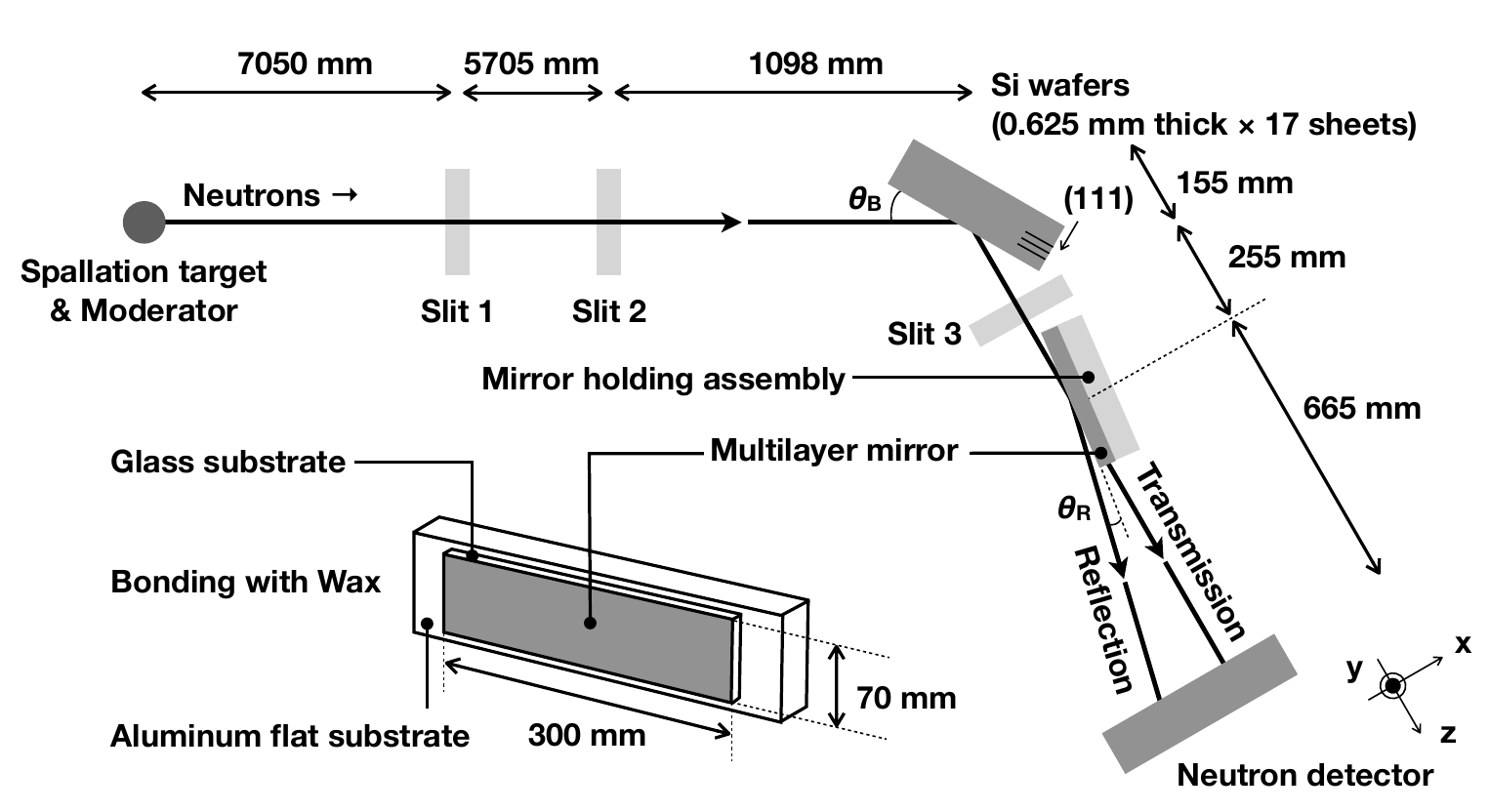}
 \caption{
    Schematic top view of the experimental setup at the J-PARC MLF BL10.
    Note that the size of the figure has not been scaled.
    The neutrons are incident on the left side of this figure.
    }
 \label{fig:setup}
\end{figure*}

The specular reflection of near-epithermal neutrons from the surface of a monochromatic NiC/Ti multilayer mirror was studied at NOBORU (BL10) of the Materials and Life Science Experimental Facility (MLF) of the Japan Proton Accelerator Research Complex (J-PARC).
Pulsed neutrons were generated through spallation reactions induced by the 3~GeV proton beam accelerated by the rapid-cycling synchrotron, with the beam incident on a mercury target\cite{Nagamiya2012-xa}.
The experiment was conducted at a repetition rate of 25~Hz and an average beam power of 700~kW. 
The neutrons were moderated by a hydrogen moderator before being supplied to each beamline.
A diagram of the experimental setup in BL10 is shown in Figure~\ref{fig:setup}. 
The supplied neutrons were collimated using a pair of 5~mm thick sintered $\rm B_4 C$ slits (slit 1 and 2), resulting in a beam width of 20~mm and a height of 80~mm. 
A mechanical chopper situated 7.2~m from the moderator removed the flame overlap of neutrons\cite{Oikawa2008-ka}.
Bragg diffraction on silicon wafers directed the collimated neutrons in different directions away from the beam axis. 
The diffracted neutrons were incident to the mirror through slit 3, which was a $50~\unit{mm}$-thick sintered $\mathrm{B_4C}$ blocks with a width of $1.13~\unit{mm}$.
The holder of slit 3 was covered with a rubber sheet containing ${}^{10}\mathrm{B}$ to absorb scattered neutrons and confine the diffracted neutron beam within a 67~mm range in the $y$-direction. 
A neutron detector based on a micro-pixel $^{3}\mathrm{He}$-containing chamber ($\mu$-PIC), capable of the recording arrival time and 2D position with a spatial resolution of $350~\unit{\mu m}$, was positioned 665~mm away from the center of the mirrors to accurately detect changes in the neutron trajectory resulting from reflections\cite{Parker2013-ib, Parker2013-cj}. 
The path length of the neutrons between the moderator and the detector was 14.93~m. 
The neutron wavelength in the pulsed neutron source was determined using time-of-flight (TOF) measurements and the distance from the moderator to the detector.

\begin{figure}[htb]
\centering
\includegraphics[keepaspectratio, width=0.5\linewidth]{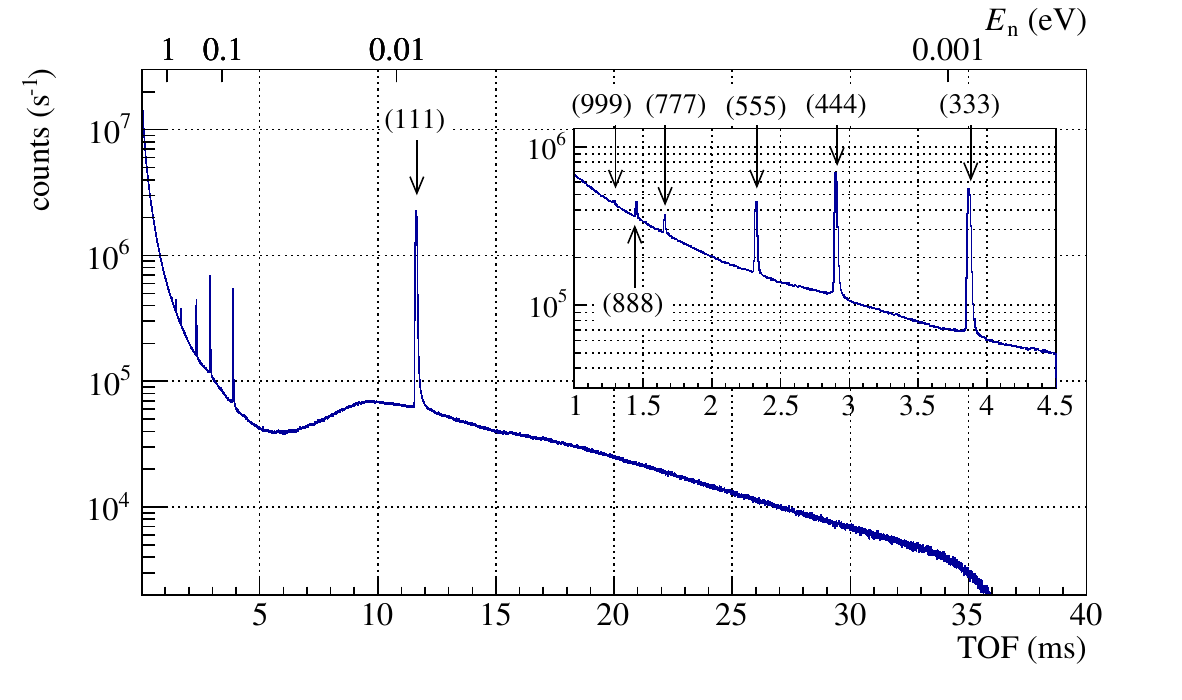}
\caption{TOF spectrum of neutrons diffracted by silicon wafers.
The inset figure shows the enlarged spectrum in the TOF region from $1~\unit{ms}$ to $4.5~\unit{ms}$.}
\label{fig:spectrum}
\end{figure}

In order to reduce the background noise of fast neutrons and flash gamma radiation originating from the neutron source, the diffracted neutrons from silicon wafers were utilized as incident neutrons for the mirror. 
We stacked 17 sheets of N-type silicon wafers, each measuring 5 inches in diameter and 0.625 mm in thickness. 
Instead of using a silicon bulk, we chose the stack of silicon wafers, as the slight misalignment between these wafers mitigates the intensity loss due to multiple diffractions.
The TOF spectrum of the diffracted neutrons was measured and depicted in Figure~\ref{fig:spectrum}. 
The intensity peak associated with a TOF around 12 ms corresponds to (111) diffraction, in accordance with the Miller index, while higher-order diffractions were identified in an earlier TOF region.
The diffraction angle $\theta_{\rm B}$ was determined to be $29.5~\unit{degrees}$, obtained from Bragg's law as $2a/\sqrt{3}\sin\theta_{\rm B} = \lambda$, where $\lambda$ represents the neutron wavelength and $a = 0.5431~\unit{nm}$ represents the lattice constant\cite{Kessler2017-up}.

\begin{figure}[htb]
\centering
\includegraphics[keepaspectratio,width=0.5\linewidth]{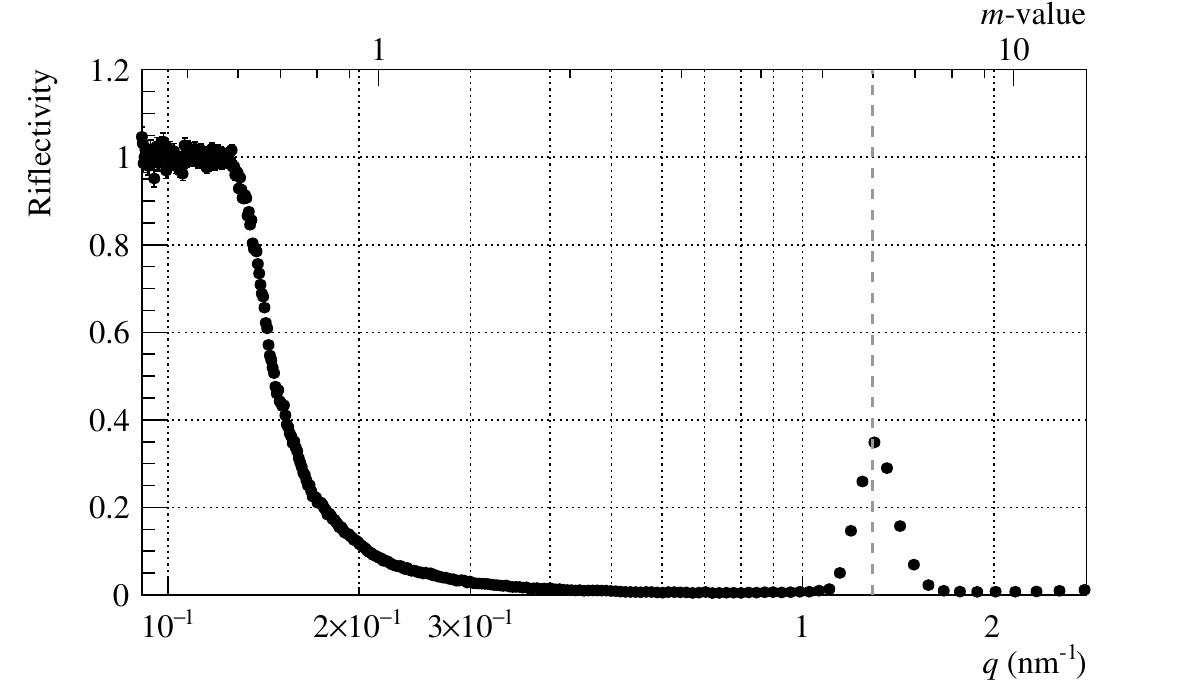}
\caption{
    The measured reflectivity of the multilayer neutron mirror used in this experiment as a function of the neutron momentum transfer $q$. 
    The dashed line represents $m=6$, where the momentum transfer is $q=1.3~\unit{nm^{-1}}$.
}
\label{fig:reflectivity}
\end{figure}

The neutron mirror employed in this study was a monochromatic NiC/Ti multilayer mirror with $m = 6$ deposited on a 3~$\unit{mm}$-thick glass substrate, $300~\unit{mm}$ long, and $70~\unit{mm}$ wide. The glass substrate with the multilayer mirror was bonded to a 30~mm-thick aluminum substrate using wax to minimize the reflection losses caused by surface waviness. 
The bonded mirror exhibited a slope error of 0.3~mrad, which was one order of magnitude smaller than the reflection angle of 0.5~eV neutrons and approximately two orders of magnitude smaller than typical neutron mirrors deposited on glass substrates. 
To compare the reflectivity of the mirror for near-epithermal and cold neutrons, the reflectivity of the mirror for cold neutrons was measured in advance at BL10. 
Reflections from the mirror were recorded using a gas electron multiplier (GEM)\cite{Uno2012-hh, Uno2012-ux} as the neutron detector capable of time and 2D position detection, which was positioned along the direct beam axis.
The reflectivity obtained as a function of the neutron momentum transfer $q$ is shown in Figure~\ref{fig:reflectivity}. 
Besides the total reflection at $q < 0.13 ~\unit{nm^{-1}}$, which is attributed to the Fermi pseudopotential of NiC/Ti, a distinct peak was observed at $q = 1.3 ~\unit{nm^{-1}}$.
This peak represents the reflection arising from Bragg diffraction of the multilayer, exhibiting a reflectivity of 35\%.
The multilayer structure was purposely designed to possess a broadened peak width, whose standard deviation was measured to be $\pm 0.09 ~\unit{nm^{-1}}$.
This broadening relaxes the angular tolerance required for aligning a monochromatic multilayer mirror to reflect a monochromatic beam, a task that is otherwise difficult.


\section{Results}

\begin{figure}[htb]
\centering
\includegraphics[keepaspectratio, width={0.5\linewidth}, height=\textheight]{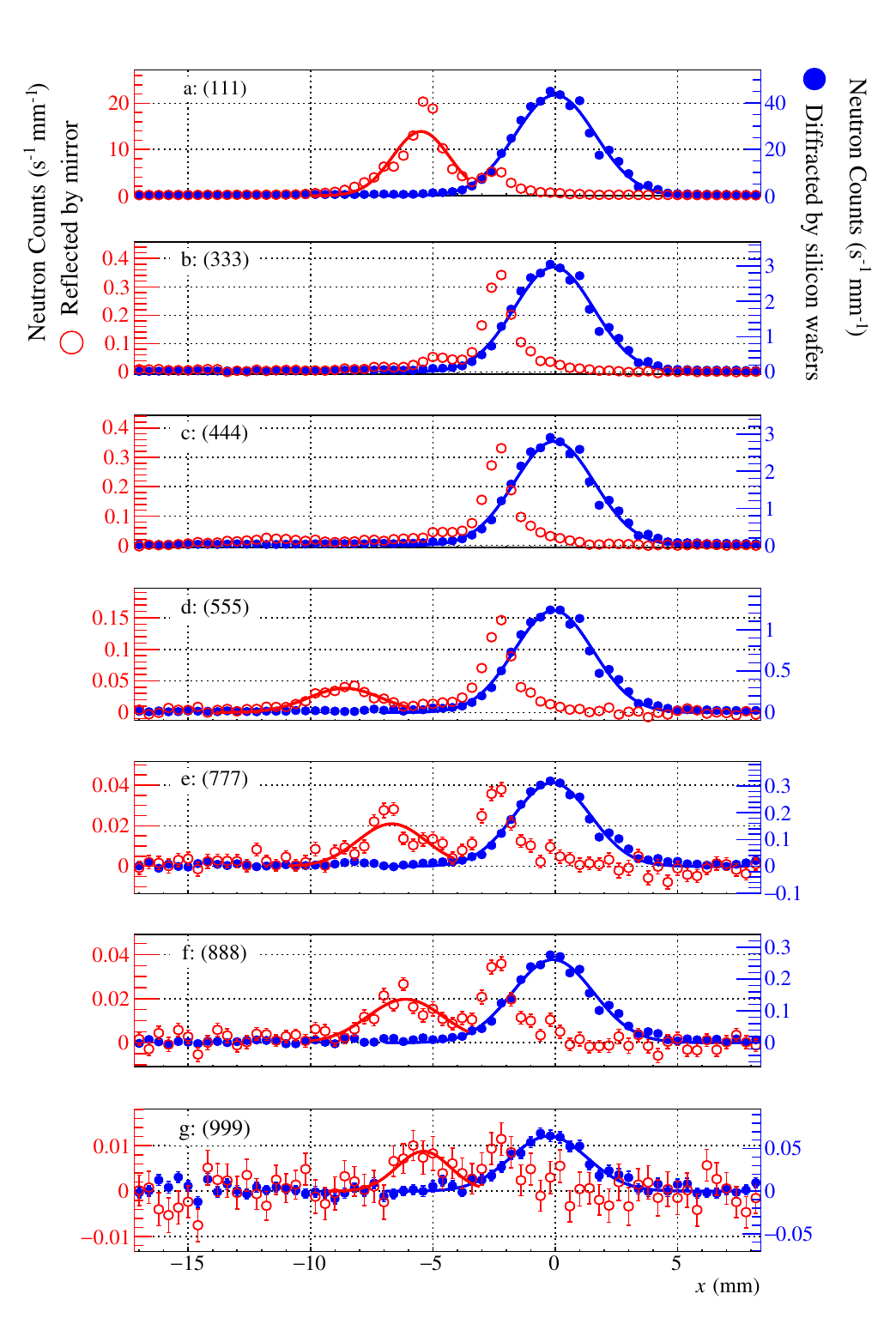}
\caption{
    (\textit{color online}). 
    The spatial distributions of the neutrons diffracted by the silicon wafers (blue, closed circle) and reflected by the mirror (red, open circle).
    The error bars represent statistical uncertainties.
    The counts of neutrons diffracted by silicon wafers are shown on the right axis, and those reflected by the mirror are shown on the left axis.
    From top to bottom, they show \lat{111} to \lat{999} diffraction, respectively.}
    \label{fig:reflection}
\end{figure}

The diffracted neutrons were incident on the mirror to investigate the near-epithermal neutron reflection. 
The spatial distributions of both reflected and diffracted neutrons are presented in Figure~\ref{fig:reflection}.
These spatial distributions were obtained by subtracting the background distributions obtained from regions where the Bragg condition was not satisfied.
We defined the position where neutrons diffracted by \lat{111} were observed as $x=0~\unit{mm}$ (see Figure~\ref{fig:reflection}a).


The diffracted neutrons from the silicon wafers exhibited peaks in spatial distributions as expected. 
The peaks of higher-order diffractions were observed at the same position as the peak of the \lat{111} diffraction, around $x=0~\unit{mm}$ (see Figure~\ref{fig:reflection}b--g). 
Diffractions by \lat{222} and \lat{666} were absent as expected because they are forbidden reflections\cite{Roberto1970-yj}. 
The highest-order diffraction observed in this study was \lat{999}.


When the multilayer mirror was inserted into the beam path, peaks indicating reflection were observed, including those in the near-epithermal region.
Specular reflection of neutrons diffracted by \lat{111} was observed at $x=-5.5~\unit{mm}$ (see Figure~\ref{fig:reflection}a).
The reflection angle at the mirror, denoted as $\theta_{\rm R}$ in Figure~\ref{fig:setup}, was determined to be $4.09~\unit{mrad}$ based on the detection position and the distance between the mirror and the detector. 
As we employed a monochromatic mirror, the reflections of diffracted neutrons from \lat{333} and \lat{444} were not observed, because they do not satisfy the momentum transfer requirements for reflection (see Figure~\ref{fig:reflection}b, c).
Notably, distinct reflection peaks of neutrons diffracted by \lat{555}, \lat{777}, \lat{888}, and \lat{999} were observed within the range of $-9<x<-5~\unit{mm}$ (see Figure~\ref{fig:reflection}e--g).
The maximum energy of the reflected neutrons diffracted by \lat{999} was determined to be 0.7~eV, thereby confirming the successful observation of specular reflection for near-epithermal neutrons.



To evaluate the reflection angle of higher energy neutrons, we introduced the positional shift ($\Delta x$) defined as the difference between the positions of the neutrons as diffracted by the silicon wafers and those reflected by the mirror. 
Gaussian fitting was applied to derive these positions from the spatial distribution data. 
Since monochromatic mirrors can reflect only neutrons that satisfy a specific momentum transfer from Figure~\ref{fig:reflectivity}, the relationship between the positional shift and the wavelength can be expressed as $\Delta x \approx L q\lambda / 2\pi$, where $q = 1.3 ~\unit{nm^{-1}}$ corresponds to the momentum transfer of the monochromatic reflection, and $L$ represents the distance between the mirror and the detector. 
The measured positional shift was $\Delta x/\lambda = 112.3 \pm 5.2 ~\unit{mm/nm}$, deviating by 24.1~mm/nm from the design value of 136.5~mm/nm. 
This discrepancy suggests that the diffracted neutrons were incident on the mirror slightly off-center due to misalignment. 
Considering the acceptable range of positional shift for the 300 mm mirror length, which falls within $\pm 30.2$~mm/nm, the measured value aligns with expectations within this range.

\begin{table*}[hbtp]
	\caption{The characteristics of neutrons diffracted by the silicon wafers.}
	\label{table:difflaction}
	\centering
     \begin{tabular}{c|ccrcr}
Order & {$E~\unit{(eV)}$} & {$q~\unit{(nm^{-1})}$} & {$R$ (\%)} & {Designed $R$ (\%)} & {$\Delta x ~\unit{(mm)}$} \\ \hline
\lat{111} & 0.00858 & 0.166 & 25.3 $\pm$ 0.0 & 25.0 & 5.5 $\pm$ 0.0 \\
\lat{333} & 0.0772 & 0.499 & 1.4 $\pm$ 0.0 & 0.9 & --- \\
\lat{444} & 0.137 & 0.666 & 2.1 $\pm$ 0.0 & 0.5 & --- \\
\lat{555} & 0.215 & 0.832 & 3.7 $\pm$ 0.1 & 0.6 & 8.5 $\pm$ 0.1 \\
\lat{777} & 0.420 & 1.17 & 7.3 $\pm$ 0.5 & 9.3 & 6.6 $\pm$ 0.1 \\
\lat{888} & 0.549 & 1.33 & 7.9 $\pm$ 0.6 & 31.5 & 6.0 $\pm$ 0.1 \\
\lat{999} & 0.695 & 1.50 & 10.7 $\pm$ 2.9 & 6.8 & 5.1 $\pm$ 0.3 \\
\end{tabular}
\end{table*}


The reflectivity ($R$) for each neutron energy was computed using the formula $R = I_{\rm ref}/\pr{I_{\rm dif} - I_{\rm gra}}$. 
Here, $I_{\rm ref}$ represents the integrated intensity of the neutrons reflected by the mirror within the region of $-14<x<-4~\unit{mm}$, $I_{\rm gra}$ corresponds to the integrated intensity of the detected neutrons with the mirror within the region of $-4<x<-1~\unit{mm}$, and $I_{\rm dif}$ denotes the integrated intensity of diffracted neutrons by the silicon wafers within the region of $-5<x<5~\unit{mm}$. 
The presence of $I_{\rm gra}$ accounts for the neutrons that were not influenced by the mirror. 
The observed reflectivity and positional shifts of each near-epithermal neutron are summarized in Table~\ref{table:difflaction}, along with the cold neutron reflectivity as Designed $R$.
The reflectivity at \lat{111} was in well agreement with the design values. 
Additionally, the reflectivities for \lat{333} and \lat{444} were close to zero, which is also consistent with the design value. 
However, for higher-order reflections beyond (444), the observed reflectivity deviated from the design values.
Notably, the reflectivity for \lat{888} was measured to be $7.9\pm0.6\%$, which was unexpectedly smaller than the design value of 31.5\%. 
The underlying cause for this discrepancy remains unidentified due to limitations in the statistical analysis. 
Improved measurements with enhanced statistics, such as utilizing large $m$-value supermirrors\cite{Neutronics_undated-gu, Eriksson2018-iw}, are necessary to gain a quantitative understanding of neutron reflectivity for practical applications.


\section*{Acknowledgments}\noindent
This research was supported by JSPS KAKENHI Grant Number 19K21876 and RIKEN Junior Research Associate Program, JST SPRING, Grant Number JPMJSP2125.
Takuhiro Fujiie and Rintaro Nakabe would like to take this opportunity to thank the ``Interdisciplinary Frontier Next-Generation Researcher Program of the Tokai Higher Education and Research System".
The neutron experiment at the Materials and Life Science Experimental Facility of the J-PARC was performed under user proposals 2020A0252, 2020B0349, and 2020B0432.
Finally, I would like to thank the support members Kenji Mishima and Masahiro Takeda.


\bibliographystyle{elsarticle-num}
\bibliography{ref}

\ifAPPEND
\clearpage
\fi

\end{document}